\documentclass[floatfix,aps,twocolumn,prl,superscriptaddress]{revtex4}
\usepackage{amsmath}
\usepackage{amssymb}
\usepackage{graphicx}
\usepackage{dcolumn}
\usepackage{natbib}
\usepackage{bm}
\usepackage{epstopdf}
\usepackage{multirow}
\begin{document}
\title{Phonon splitting and anomalous enhancement of infrared-active modes in BaFe$_{2}$As$_{2}$}
\author{A. A. Schafgans}
\email{aschafgans@physics.ucsd.edu}
\affiliation{Department of Physics, University of California, San Diego, La Jolla, California 92093, USA}
\author{B. C. Pursley}
\affiliation{Department of Physics, University of California, San Diego, La Jolla, California 92093, USA}
\author{A. D. LaForge}
\affiliation{Department of Physics, University of California, San Diego, La Jolla, California 92093, USA}
\author{A. S. Sefat}
\affiliation{Materials Science and Technology Division, Oak Ridge National Laboratory, Oak Ridge, Tennessee 37831, USA}
\author{D. Mandrus}
\affiliation{Materials Science and Technology Division, Oak Ridge National Laboratory, Oak Ridge, Tennessee 37831, USA}
\affiliation{Department of Materials Science and Engineering, University of Tennessee, Knoxville, TN 37996}
\author{D. N. Basov}
\affiliation{Department of Physics, University of California, San Diego, La Jolla, California 92093, USA}
\date{\today}

\begin{abstract}
We present a comprehensive infrared spectroscopic study of lattice dynamics in the pnictide parent compound BaFe$_{2}$As$_{2}$. In the tetragonal structural phase, we observe the two degenerate symmetry-allowed in-plane infrared active phonon modes. Following the structural transition from the tetragonal to orthorhombic phase, we observe splitting into four non-degenerate phonon modes and a significant phonon strength enhancement. These detailed data allow us to provide a physical explanation for the anomalous phonon strength enhancement as the result of anisotropic conductivity due to Hund's coupling.
\end{abstract}

\maketitle

The pnictide high temperature superconductors display a rich phase diagram including temperature and doping dependent structural and magnetic phase transitions in proximity to the superconducting phase \cite{Paglione-NatPhys6-645-2010,Basov-Chubukov,Basov-RMP}. Such complexity has myriad observational consequences. Phonon behavior provides a unique window into the phase diagram of the pnictides and as we will show, reveals a fascinating interplay between structure, charge and magnetism. Infrared (IR) spectroscopic studies consistently showed an anomalous phonon strength enhancement at low temperatures in the 122 and 1111 families  \cite{Akrap-preprint,Dong-PRB82-054522-2010,Martini-EPL84-67013-2008,WuPRB}, yet these studies did not observe all of the phonon modes predicted by group theory. In order to understand the origin of the phonon strength enhancement, it is necessary to experimentally distinguish each phonon mode in the orthorhombic phase. In this work, by observing all of the predicted phonon modes at low temperature, we are able to comment on the origins of the phonon strength enhancement.

The samples in this study were square platelet single crystals of BaFe$_{2}$As$_{2}$ (Ba122) approximately 2 x 2 x 0.1 mm in size \cite{Sefat-arXiv:0807.2237}. In Ba122, two phase transitions have been observed as a function of temperature; a structural phase transition from high temperature tetragonal (HTT, TrCr$_{2}$Si$_{2}$ type) to low temperature orthorhombic (LTO) at $T_{STR}\approx$140K, and a magnetic phase transition from paramagnetic (PM) order to spin density wave (SDW) order below $T_{SDW} \approx T_{STR} \approx$ 140 K \cite{Sefat-arXiv:0807.2237,Hu-PRL101-257005-2008,Sefat-PRB79-094508-2009}. Fabrication and characterization are described elsewhere \cite{Sefat-arXiv:0807.2237}. 

We measured near-normal incidence reflectance \emph{R}($\omega$) of the \emph{ab} face, over a frequency range of $\omega \approx$ 20 to 12,000 cm$^{-1}$. The reflectance measurements were performed for a variety of temperatures ranging from \emph{T} = 6.5 K to 295 K. Additionally, we performed variable-angle spectroscopic ellipsometry from $\omega \approx$ 5,500 to 45,000 cm$^{-1}$. In order to extract the optical constants, we performed a Kramers-Kronig constrained variational analysis using refFIT software, as detailed in the references \cite{Kuzmenko-RevSciInstrum76-083108-2005,Qazilbash-arXiv:0808.3748}.
\begin{figure}
\centering
\includegraphics[width=3.5in]{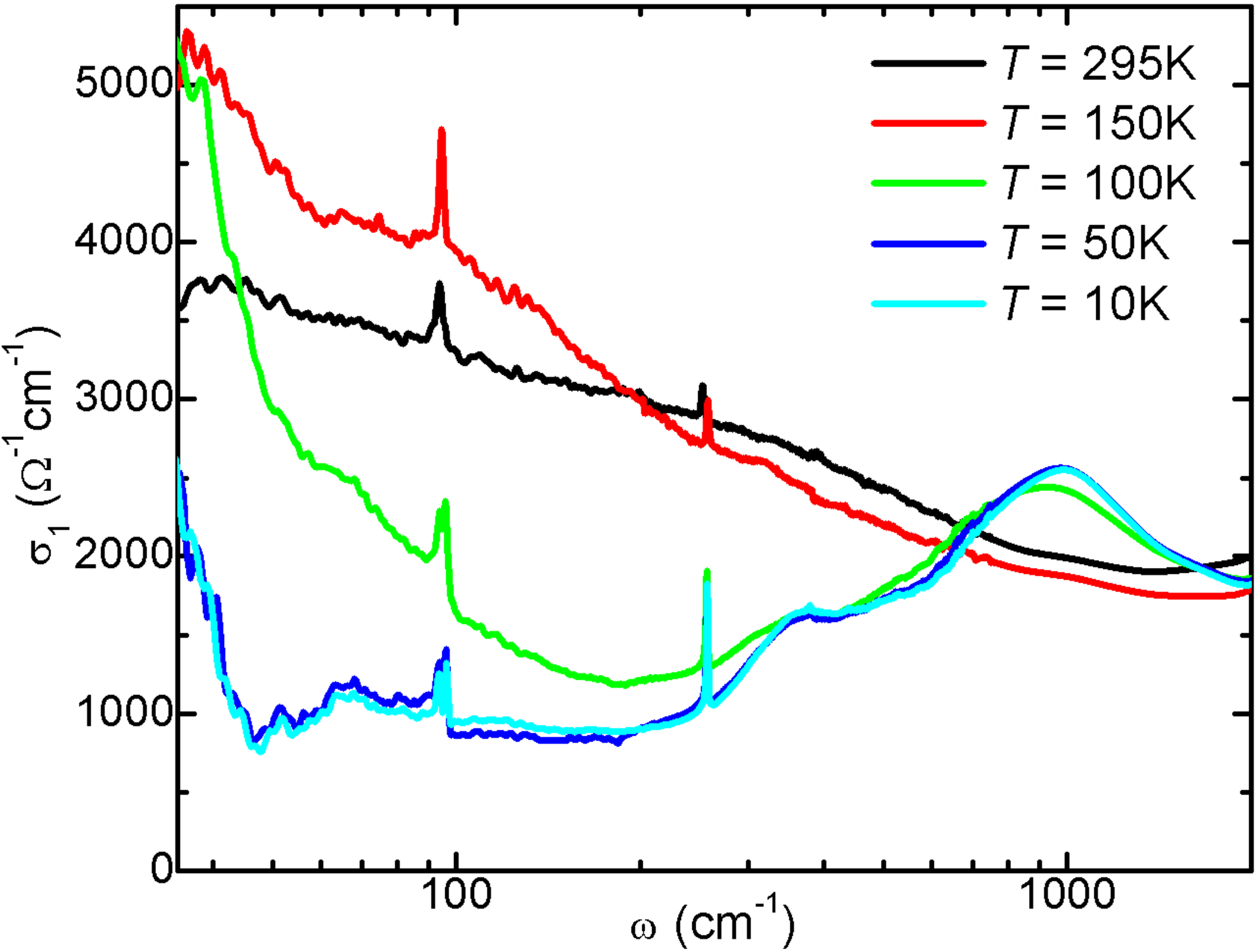}
\caption{Temperature dependence of the real part of the optical conductivity $\sigma_1(\omega)$ in Ba122, focused on the far-infrared frequency range.}
\label{Fig. 2}
\end{figure}

The optical conductivity $\sigma_{1}(\omega)$ is plotted in Figure 1. At low energy, a conspicuous Drude response is present for $T>T_{SDW}$, becoming mostly gapped at low temperature. For temperatures $T < T_{SDW}$, a significant loss of conductivity below $\approx$ 700 cm$^{-1}$ results in the development of a large optical transition centered at $\approx$ 1,000 cm$^{-1}$ where $\sigma_{1}$ is greater than the PM state value extending to 2000 cm$^{-1}$. Additionally, there is a smaller optical transition that develops near 350 cm$^{-1}$. As has been done previously \cite{Moon-PRB81-204114-2010,Hu-PRL101-257005-2008}, we attribute the onset of these optical transitions to the redistribution of spectral weight from the region of the SDW gap. We determined the two SDW gap values in another work \cite{Schafgans-unpub}: $E_{g}^{1}$ = 336 $\pm$ 3 cm$^{-1}$ and $E_{g}^{2}$ = 656 $\pm$ 2 cm$^{-1}$. Two phonon modes appear in the conductivity and are located below the lower gap energy. These phonons will be the focus of the remainder of this manuscript.
\begin{figure*}
\centering
\includegraphics[width=7in]{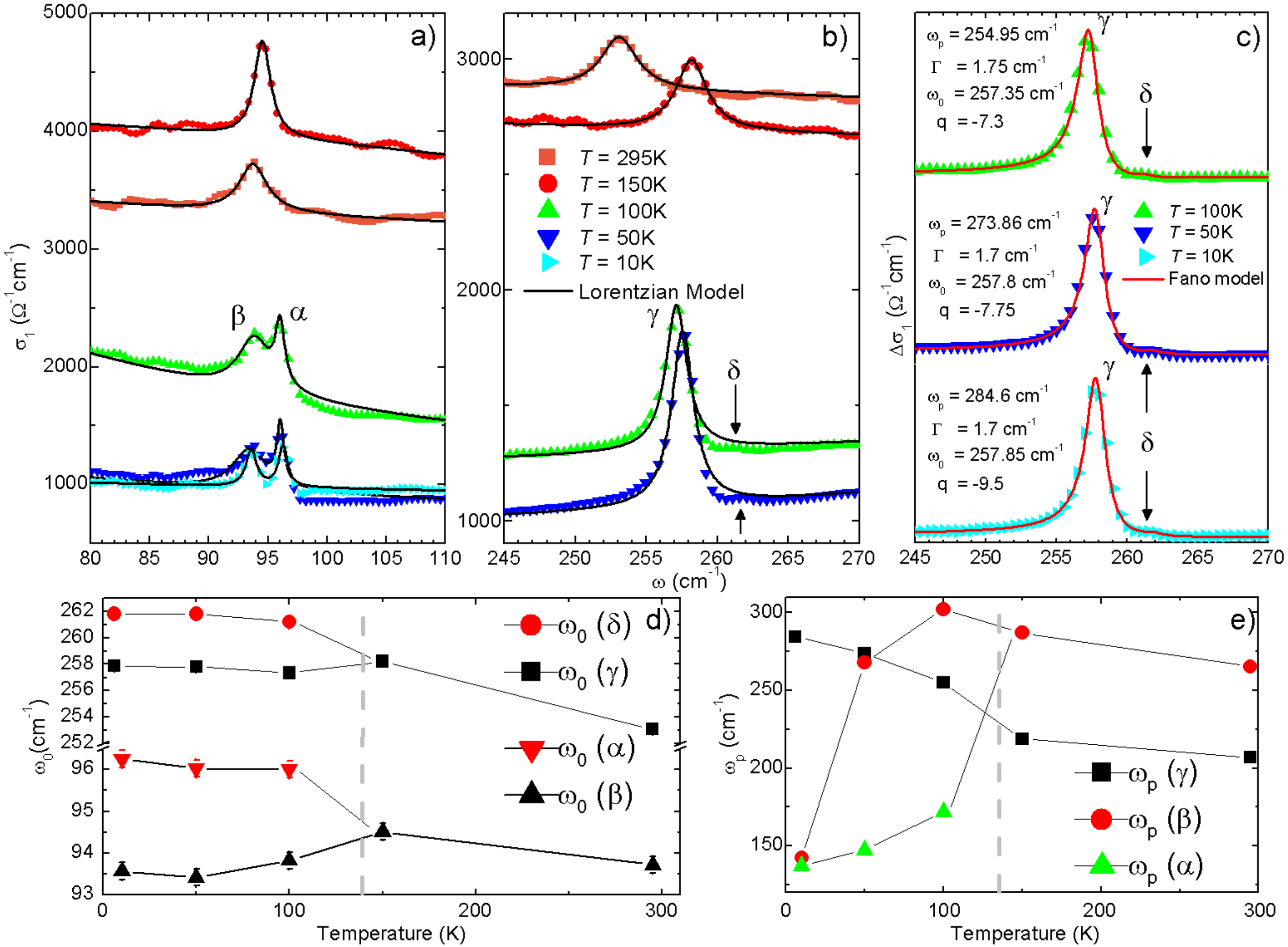}
\caption{The real part of the optical conductivity $\sigma_1(\omega)$, focusing on the phonon modes in Ba122 near 95 $cm^{-1}$ (Ba-Ba, panel a) and 257 cm$^{-1}$ (Fe-As, panels b and c). We observe both phonon modes to split upon the transition from the HTT to LTO phases. Panels a and b show Lorenzian fits to the data, while panel c shows fits of the Fano form (Eq. 1) to the phonon at 257 cm$^{-1}$. The fit parameters are shown graphically in panel d (oscillator frequency position, $\omega_0$) and panel e (oscillator strength, $\omega_p$). Table I summarizes the results of our fits.}
\label{Fig. 2}
\end{figure*}

In the HTT phase, Ba122 is predicted to exhibit two symmetry-allowed \emph{ab} plane IR-active \emph{E}$_{u}$ modes \cite{Litvinchuk-PRB78-060503-2008} (Figs. 2a and b). We observe both modes, almost identical to the recent report by Akrap, \emph{et. al.} \cite{Akrap-preprint}: one at 94.5 cm$^{-1}$ due to Ba(\emph{ab}) displacements and the other at 254.1 cm$^{-1}$ due to Fe(\emph{ab}) and As(\emph{-ab}) displacements. The 254.1 cm$^{-1}$ phonon has been observed in many of the \emph{A}Fe$_{2}$As$_{2}$ materials, including \emph{A} = Ca, Sr, Eu, and Ba \cite{Mittal-arxiv09024590,Litvinchuk-PRB78-060503-2008,Hu-arxiv09020435,Mittal-PRB78-104514-2008}, whereas the position of the phonon due to the alkali element \emph{A} varies more widely. Upon the structural transition to the LTO phase below $T_{STR}$, the degeneracy of the two \emph{E}$_{u}$ modes is broken because of the unequal \emph{a} and \emph{b}-axis bond lengths, becoming $B_{2u}+B_{3u}$ along the \emph{b} and \emph{a} axes, respectively \cite{Akrap-preprint}. Therefore, phonon splitting is expected to occur \cite{Litvinchuk-PRB78-060503-2008}, and four IR-active modes should become evident. In Figure 2, we show the unambiguous observation of phonon splitting using IR optics. We fit the phonons using Lorentzian oscillators, the results of which are given in Table 1 and plotted in Fig. 2. (We note that phonon splitting has been observed using Raman spectroscopy \cite{Raman}.)

\begin{table*}
\caption{The oscillator position $\omega_0$, oscillator strength $\omega_p$, and oscillator width $\Gamma$, based on the models used to fit the optical spectra. The error for each fit value is listed in parentheses. For the Fano oscillators, the Fano parameter is listed in Figure 2c.}
\begin{center}
\begin{tabular}{c | c | c | c | c | c | c}
\cline{2-6}
& \multicolumn{5}{| c |}{Ba modes} & \\
\cline{2-6}
& \multicolumn{2}{| c |}{HTT} & \multicolumn{3}{| c |}{LTO} & \\
\cline{2-6}
& \multicolumn{1}{| c |}{295K} & \multicolumn{1}{| c |}{150K} & \multicolumn{1}{| c |}{100K} & \multicolumn{1}{| c |}{50K} & \multicolumn{1}{| c |}{10K} & \\
\cline{1-7}
\multicolumn{1}{| c |}{\multirow{2}{*}{$\omega_0$}} & \multicolumn{1}{| c |}{\multirow{2}{*}{93.7(0.3)}} & \multicolumn{1}{| c |}{\multirow{2}{*}{94.5(0.3)}} & \multicolumn{1}{| c |}{93.8(0.3)} & \multicolumn{1}{| c |}{93.4(0.3)} & \multicolumn{1}{| c |}{93.6(0.3)} & \multicolumn{1}{| c |}{$\alpha$} \\ 
\cline{4-7}
\multicolumn{1}{| c |}{} & \multicolumn{1}{| c |}{} & \multicolumn{1}{| c |}{} & \multicolumn{1}{| c |}{96.0(0.2)} & \multicolumn{1}{| c |}{96.0(0.2)} & \multicolumn{1}{| c |}{96.3(0.2)} & \multicolumn{1}{| c |}{$\beta$} \\  
\cline{1-7}
\multicolumn{1}{| c |}{\multirow{2}{*}{$\omega_p$}} & \multicolumn{1}{| c |}{\multirow{2}{*}{265.3(10)}} & \multicolumn{1}{| c |}{\multirow{2}{*}{287.0(10)}} & \multicolumn{1}{| c |}{302.3(10)} & \multicolumn{1}{| c |}{267.9(10)} & \multicolumn{1}{| c |}{142.3(10)} & \multicolumn{1}{| c |}{$\alpha$} \\ 
\cline{4-7}
\multicolumn{1}{| c |}{} & \multicolumn{1}{| c |}{} & \multicolumn{1}{| c |}{} & \multicolumn{1}{| c |}{171.7(10)} & \multicolumn{1}{| c |}{147.52(10)} & \multicolumn{1}{| c |}{137.3(10)} & \multicolumn{1}{| c |}{$\beta$} \\  
\cline{1-7}
\multicolumn{1}{| c |}{\multirow{2}{*}{$\Gamma$}} & \multicolumn{1}{| c |}{\multirow{2}{*}{2.9(0.2)}} & \multicolumn{1}{| c |}{\multirow{2}{*}{1.7(0.1)}} & \multicolumn{1}{| c |}{3.4(0.2)} & \multicolumn{1}{| c |}{3.5(0.2)} & \multicolumn{1}{| c |}{1.1(0.1)} & \multicolumn{1}{| c |}{$\alpha$} \\ 
\cline{4-7}
\multicolumn{1}{| c |}{} & \multicolumn{1}{| c |}{} & \multicolumn{1}{| c |}{} & \multicolumn{1}{| c |}{0.9(0.1)} & \multicolumn{1}{| c |}{0.7(0.1)} & \multicolumn{1}{| c |}{0.9(0.1)} & \multicolumn{1}{| c |}{$\beta$} \\  
\cline{1-7}
\end{tabular}
\end{center}

\begin{center}
\begin{tabular}{c | c | c | c | c | c | c}
\cline{2-6}
& \multicolumn{5}{| c |}{Fe-As modes} & \\
\cline{2-6}
& \multicolumn{2}{| c |}{HTT} & \multicolumn{3}{| c |}{LTO} & \\
\cline{2-6}
& \multicolumn{1}{| c |}{295K} & \multicolumn{1}{| c |}{150K} & \multicolumn{1}{| c |}{100K} & \multicolumn{1}{| c |}{50K} & \multicolumn{1}{| c |}{10K} & \\
\cline{1-7}
\multicolumn{1}{| c |}{\multirow{2}{*}{$\omega_0$}} & \multicolumn{1}{| c |}{\multirow{2}{*}{253.1(0.3)}} & \multicolumn{1}{| c |}{\multirow{2}{*}{258.2(0.3)}} & \multicolumn{1}{| c |}{257.4(0.3)} & \multicolumn{1}{| c |}{257.8(0.3)} & \multicolumn{1}{| c |}{257.9(0.3)} & \multicolumn{1}{| c |}{$\gamma$} \\ 
\cline{4-7}
\multicolumn{1}{| c |}{} & \multicolumn{1}{| c |}{} & \multicolumn{1}{| c |}{} & \multicolumn{1}{| c |}{261.2(0.2)} & \multicolumn{1}{| c |}{261.8(0.2)} & \multicolumn{1}{| c |}{261.8(0.2)} & \multicolumn{1}{| c |}{$\delta$} \\  
\cline{1-7}
\multicolumn{1}{| c |}{\multirow{2}{*}{$\omega_p$}} & \multicolumn{1}{| c |}{\multirow{2}{*}{207.0(10)}} & \multicolumn{1}{| c |}{\multirow{2}{*}{218.8(10)}} & \multicolumn{1}{| c |}{255.0(10)} & \multicolumn{1}{| c |}{273.9(10)} & \multicolumn{1}{| c |}{284.6(10)} & \multicolumn{1}{| c |}{$\gamma$} \\ 
\cline{4-7}
\multicolumn{1}{| c |}{} & \multicolumn{1}{| c |}{} & \multicolumn{1}{| c |}{} & \multicolumn{1}{| c |}{31.6(7)} & \multicolumn{1}{| c |}{38.7(7)} & \multicolumn{1}{| c |}{38.7(7)} & \multicolumn{1}{| c |}{$\delta$} \\  
\cline{1-7}
\multicolumn{1}{| c |}{\multirow{2}{*}{$\Gamma$}} & \multicolumn{1}{| c |}{\multirow{2}{*}{3.1(0.4)}} & \multicolumn{1}{| c |}{\multirow{2}{*}{2.6(0.4)}} & \multicolumn{1}{| c |}{1.8(0.2)} & \multicolumn{1}{| c |}{1.7(0.2)} & \multicolumn{1}{| c |}{1.7(0.1)} & \multicolumn{1}{| c |}{$\gamma$} \\ 
\cline{4-7}
\multicolumn{1}{| c |}{} & \multicolumn{1}{| c |}{} & \multicolumn{1}{| c |}{} & \multicolumn{1}{| c |}{1.5(0.5)} & \multicolumn{1}{| c |}{1.5(0.5)} & \multicolumn{1}{| c |}{1.5(0.5)} & \multicolumn{1}{| c |}{$\delta$} \\  
\cline{1-7}
\end{tabular}
\end{center}
\end{table*}

Figure 2a shows the temperature dependence of the conductivity, focusing on the lower energy phonon at $\omega_{0} \approx$ 94.5 cm$^{-1}$. The mode hardens by 1 cm$^{-1}$ upon cooling to $T_{STR}$, and then it splits into two distinguishable peaks. The two modes are separated by 2.9 cm$^{-1}$, with the lower phonon ($\beta$) centered at the original room temperature position of $\omega_{0}$ = 94.5 cm$^{-1}$. Identifying the Ba-Ba distance with the \emph{a}=5.6146$\AA$ and \emph{b}=5.5742$\AA$ lattice constants in the LTO phase \cite{Rotter-PRB78-020503R-2008}, we can use a simple equation to estimate the expected splitting of the phonon frequency $\omega^{\alpha}_{0}/\omega^{\beta}_{0} = (\emph{l}_{\beta}/\emph{l}_{\alpha})^{(3/2)}$, where \emph{l} is the bond length \cite{Hadjiev-PRB77-220505-2008}. This yields an expected splitting of 1.1\% (1 cm$^{-1}$); much smaller than the observed value of 2.9 cm$^{-1}$. We do not observe any temperature dependence of the position ($\omega_0$) of these two modes below $T_{STR}$. In the LTO phase, just below $T_{STR}$, the combined oscillator strength of the $\alpha$ and $\beta$ modes ($\omega_p$) is almost twice the HTT value. After the initial increase, the combined spectral weight is greatly reduced by \emph{T} = 10K. This strange behavior of the combined spectral weight may be due to the width of the modes becoming resolution limited at low temperature. However, to our knowledge, no previous report has found any change in the spectral weight of the Ba-Ba mode across the structural phase transition. This may be due to the fact that splitting of this mode has not previously been reported.

Below $T_{STR}$, the Fe-As mode (Fig. 2b) gains substantial strength. Our observations indicate that both the 94.5 cm$^{-1}$ and 254.1 cm$^{-1}$ modes experience significantly enhanced total oscillator strength in the LTO phase. We observe evidence of the Fe-As phonon splitting, with a weak higher energy phonon ($\delta$ mode) at 261.8 cm$^{-1}$. Identifying the Fe-Fe bond length as 2.808 and 2.787 $\AA$ \cite{Rotter-PRB78-020503R-2008}, the expected orthorhombic splitting of 1.13\% is close to the observed splitting of 1.5\% $\pm$ 0.2\% (3.9 cm$^{-1}$). The fact that the $\gamma$-mode moves downward in energy below $T_{STR}$ while the $\delta$-mode remains at higher energy na\"{i}vely implies that the $\gamma$-mode corresponds to the longer of the two bond lengths. As we will discuss later, the $\gamma$-mode is in fact observed to result from the shorter \emph{b}-axis bond length. Additionally, we performed similar measurements in the Co-doped superconducting compound BaFe$_{1.84}$Co$_{0.16}$As$_{2}$. There is no phonon strength enhancement at low temperatures in the doped compound (Figure 3), supporting the notion that the particular properties of the LTO and SDW phases are required to produce the phonon strength enhancement.

Lattice phonon modes generally result in a Lorentzian lineshape, symmetric about a central frequency. Coupling of the phonon to a continuum of charge and spin excitations produces an asymmetric lineshape. The theoretical foundation for describing such coupling was developed initially by Fano \cite{Fano}. Written in terms of the optical conductivity \cite{Kuzmenko-PRL103-116804-2009}, the Fano oscillator is a modified form of the standard Lorentz oscillator:
\begin{equation}
{{\sigma_1(\omega)} = {{\omega^{2}_{p}\over{60\Gamma}}}{{{q^2+{2q(\omega-\omega_0)\over{\Gamma}}}-1}\over{{q^2(1+{4(\omega-\omega_0)^2\over{\Gamma^2}}}})}},
\end{equation}
where \emph{q} is the Fano parameter which provides asymmetry, $\omega_p$ is the oscillator stength, $\omega_0$ the central frequency, and $\Gamma$ is the broadening of the oscillator. We determined that a Lorentzian oscillator was insufficient to accurately describe the lineshape of the 258 cm$^{-1}$ phonon in the SDW state and instead modeled this mode using the Fano oscillator (Fig. 2c). We find a small but measurable asymmetry, resulting in a modest Fano parameter value, which decreases as the temperature is lowered. The decrease in asymmetry can be understood by considering that in the gapped state, the coherent electronic states are continuously reduced as the temperature is lowered. Such depletion of the coherent quasiparticles results in a smaller coupling of the phonon mode to the coherent electronic states. Yet, as \emph{q} grows in magnitude and the lineshape becomes more symmetric, we do not see a corresponding decrease in the mode intensity. After considering many scenarios for both phonon strength enhancement and coupling with the electronic background that decreases at the lowest temperatures, we have identified only one possible explanation.
\begin{figure}
\centering
\includegraphics[width=3.4in]{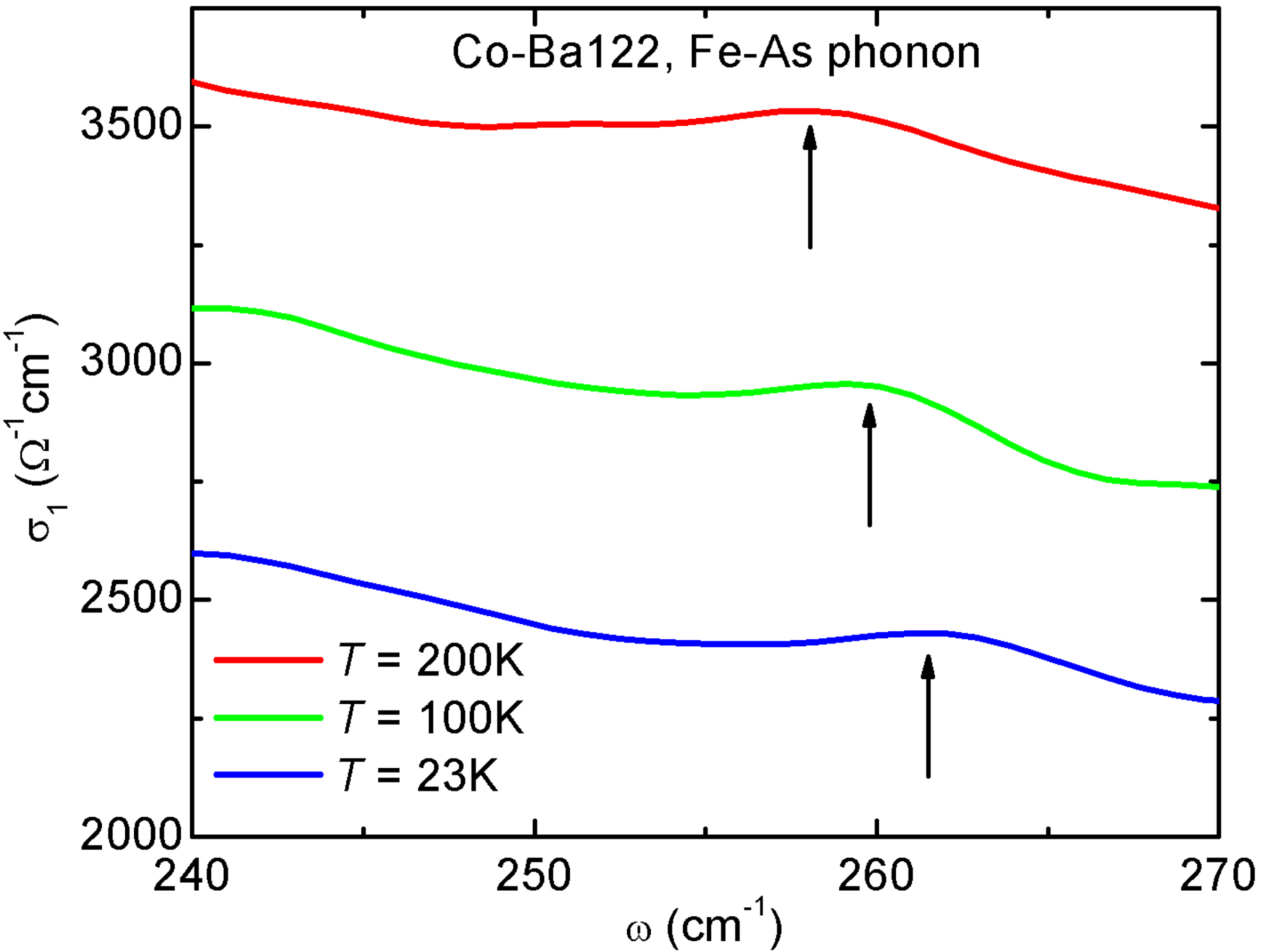}
\caption{Temperature evolution of the 257 cm$^{-1}$ phonon mode in BaFe$_{1.84}$Co$_{0.16}$As$_{2}$. We find this mode to evolve monotonically with temperature and to be just above the noise floor of our measurement for temperatures at 200K and below.}
\label{Fig. 3}
\end{figure}

Hund's rule coupling has been shown to be very important for understanding the strong correlations in the pnictides \cite{Yin-arXiv1007.2867,Schafgans-unpub,Wang-arxiv}. In addition to correlations, Hund's coupling results in very anisotropic electronic conductivity (Drude response) \cite{Yin-arXiv1007.2867,Chu-Science329-824-2010}. Recent polarized infrared studies of detwinned Ba122 crystals \cite{Degiorgi-arxiv10072543,Dusza,Uchida-arXiv} have observed drastically anisotropic optical conductivity between the \emph{a} and \emph{b} axes. The Drude response is significantly reduced and the Fe-As phonon enhanced for \emph{b}-axis polarization while for \emph{a}-axis polarization, the Drude response remains much stronger and the Fe-As phonon is diminished \cite{Uchida-arXiv}. It has been shown \cite{Kofu-arXiv0901.0738} that both the magnetic moment and the anitferromagnetic wavevector are along the \emph{a}-axis while the ferromagnetic wavevector aligns with the \emph{b}-axis. The Pauli exclusion principle favors conductance along the antiferromagnetic direction, and therefore the primary conductance channel is aligned with the \emph{a}-axis while conductance along the \emph{b}-axis is suppressed \cite{Yin-arXiv1007.2867,Chu-Science329-824-2010,Uchida-arXiv}.

For a phonon coupled to the free electron response in an orthorhombic crystal such as Ba122, anisotropic conductivity will result in additional screening of the mode in the direction of preferred conductance, leading to a reduction in the oscillator strength. Along the suppressed conductivity direction, one would expect an enhancement in the phonon strength due to the lower electronic screening. Furthermore, the phonon should become more symmetric at the lowest temperatures as the continuum of Drude states are progressively gapped by the SDW order. This is precisely the situation we observe. Therefore, we conclude that the anomalies of the spectral weight of the $\gamma$- and $\delta$-modes is caused by anisotropic conductivity due to Hund's rule coupling. Our results uncover yet another anomaly: since the enhanced phonon is polarized along the shorter \emph{b}-axis, one would expect its resonant frequency to be higher than the phonon polarized along the longer \emph{a}-axis. The observed frequency dependence of the phonon modes is a reversal from expectations. Theoretical studies are necessary to directly address the consequences of Hund's coupling on lattice dynamics. Finally, we note that with small amounts of substitutional Co-doping, the DC conductivity anisotropy becomes enhanced \cite{Chu-Science329-824-2010}, but the effect upon the phonon strength enhancement is unclear. We have observed that the Fe-As phonon is almost completely washed out with \emph{x}=0.16 Co-doping, but an IR study of the intermediate doping regime where the LTO, SDW, and superconducting phases coexist will be crucial to address this open question.

This work was supported by the NSF 1005493 and the AFOSR. D. Mandrus was supported by the U.S. Department of Energy, Basic Energy Sciences, Materials Sciences and Engineering (AS).

\end{document}